# LANSCE 805 MHZ LOW LEVEL RF RESONANCE CONTROL WATER SYSTEM UPGRADE

A. Poudel†, J. Medina, J. O'Hara, H. Salazar
Los Alamos National Laboratory, Los Alamos, NM, USA
J. Montross, Oak Ridge National Laboratory, Oak Ridge, TN, USA

*Abstract*

The Los Alamos Neutron Science Center (LANSCE) accelerator at Los Alamos National Laboratory (LANL) has been in service for over 50 years. Efforts to update and modernize crucial systems, many of which are original, are ongoing. This paper reports on the refurbishment of the Low-Level Radio Frequency (LLRF) Resonance Control Water System (RCWS) for the half-mile long Cavity-Coupled LINAC (CCL). The RCWS controls the resonance frequency of the cavities by controlling the temperature of the cooling water delivered to each of 44 accelerator modules. Of the 44 modules making up the CCL, 20 now have an upgraded RCWS. Upgrading includes removing the old hardware and installing new components including: water pumps, mix tanks, valves, temperature switches, flow switches and plumbing. This paper describes the design of the new system, material and component selection, installation, and technical challenges.

## INTRODUCTION

The resonant frequency of the CCL accelerator cavities depends sensitively on their physical dimensions, which because of thermal expansion and contraction are in turn a function of the structure temperature. The frequency decreases when a warming cavity expands. As the cavity cools, the frequency increases. The resonant frequency of the cavities must be maintained constant at the frequency of the input power in order for the maximum power to be transferred to the structure [1]. The resonance control water system tunes the temperature of the cavities by controlling the temperature of the cooling water delivered to the module in which they are housed.

## OVERVIEW OF 805 MHZ LLRF RCWS

Each of the 44 modules making up the LANSCE CCL has a dedicated RCWS. It consists of a mix tank, accelerator cooling water main supply and return manifolds, manually operated isolation valves, resonance control (RC) valve which is an automated valve that regulates the temperature of the cooling water delivered to the module, balancing valve which is a manual valve to control the amount of hot water leaving the mix tank and flowing to return manifold, a booster pump, an automated pump that maintains the flow/pressure on the cooling loop, temperature switches on the mix tank and booster pump to monitor for overheating, and flow switches for monitoring the water flow rate on the return manifolds. Detection of temperature or flow rate deviations by temperature or flow switches turn off the booster pumps which turns off the RF power.

___________________
* Work supported by the United States Department of Energy, NNSA.
† anju@lanl.gov

It then drops the accelerator "Run Permit" and shuts down beam production.

The cooling system for all the modules is identical in functionality. Cool water from the main supply passes through the resonance control valve into the mix tank. It mixes with the warm water returning from the accelerating structure which is then pressurized by the booster pump and delivered to the accelerating structures via supply header. Warm water returning from the return header is then delivered back to the mix tank and mixes with the cool water flowing into the mixing tank. A portion of the mixed water is bled off the mix tank through the balancing valve and returned to the main return manifold and to the heat exchanger outside the tunnel.

The 805 MHz temperature control system maintains the resonance of an RF module by controlling the temperature of water circulating through cooling passages attached around the structure. In order to maintain evenly distributed cooling throughout the length of the module, the supply water to the cooling passages is split and distributed to each end of the module. As seen in Figure 1, there exists an inlet and outlet at each end of the module. Water temperature is sensed at each of these points by means of four thermistors and fed back to the control electronics. T2 and T4 are thermistors for inlet while T1 and T3 are for outlet. As the average power increases, heat dissipated in the structure increases, so the water temperature must decrease to maintain resonance while removing additional heat. The control electronics drive the resonance control valve and demands more cooling water as the temperature of the structure goes up with the increase in average power.

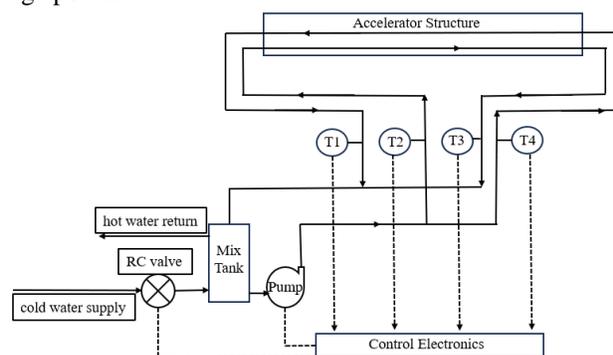

Figure 1: 805 MHz LLRF RCWS Temperature Control

Prior to the beginning of the RCWS upgrade campaign in 2021, most of the pumps, mix tanks, valves, and plumbing were original parts from construction of LANSCE in the 1960s. Dynamic instabilities were introduced by the degraded state of the components, especially the valves. As seen in Figure 2, copper tubing was used, which is a sub-optimal material for carrying de-ionized (DI) cooling water. The resulting solid debris accumulated and caused partial blockages, leading to beam down time and increased personnel entries into beam tunnels. Moreover, it is likely the wall thickness of the tubing has decreased after fifty years of DI service.

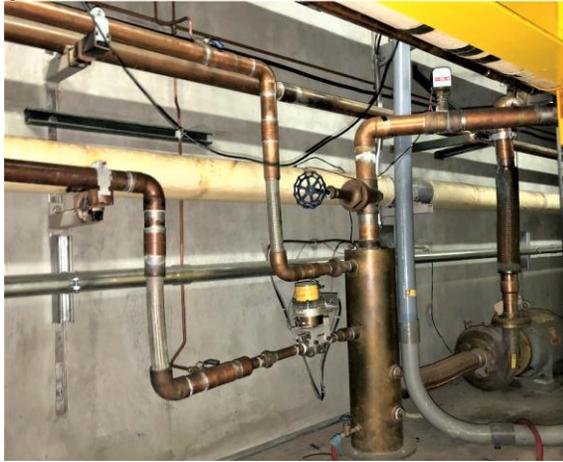

Figure 2: Original 805 MHz LLRF RCWS

## DESIGN, FABRICATION AND MATERIAL SELECTION

Given the critical function the RCWS provides, risk was an important input to early decisions made about scope of the upgrade project. The likelihood of upgraded RCWSs working immediately with minimal commissioning had to be maximized. Therefore, the original 1960s design and layout of the system was preserved. Changes were limited to materials, and the upgrade would replace booster pumps, their isolation pads, mix tanks, valves, and instrumentation. Most of the plumbing and parts, such as the pipes, valves, fittings, mix tanks, etc. were fabricated from 316 stainless steel (SS). These materials were chosen for their compatibility with DI water, resistance to radiation effects, ease of manufacturing and costs. Brass flow switches were also replaced with stainless steel ones. The high accuracy ($\pm 1°C$), Klixon ® 4344 series temperature switches were designed and fabricated by Flame Enterprises Inc. [2] to be used on the mix tanks and booster pumps.

Components such as pipe, valve, and pump specifications were sized to mimic the original assembly. This was done to avoid new operational risks. Old documents were heavily consulted, and hand calculations were performed to verify sizing and specifications of the existing systems. The characteristics of the 805 MHz LLRF RCWS are shown in Table 1.

SS Victaulic fittings were chosen for piping connections due to faster installation times, easier maintainability, reduced downtime on system retrofits, and compliance with the welding prohibition in the beam tunnel. Not all required fittings were available in SS, therefore some reducers and mechanical tees made from carbon steel were used. It was verified that the carbon steel would not be in contact with DI water. EPDM and viton seals were used in the fittings due to their higher resistance to radiation effects.

The mix tanks were newly designed to use stainless steel while matching the specifications of the original copper ones. Finite element analysis was performed using SolidWorks to verify conformance with American Society of Mechanical Engineering (ASME) Boiler and Pressure Vessel Code [3]. Terminal Engineering & Manufacturing [4] fabricated twenty-eight mix tanks. The mix tanks were hydrostatically pressure tested to withstand 1.0 MPa (150 psi) at 38 ℃ (100 °F) temperature.

The Pentair 3801 booster pump was selected to replace the original 1960s AURORA GBHA-AB. The 11.2 kW (15 hp) Pentair is rated at 1136 L m$^{-1}$ (300 GPM) for a pressure rise of 269 kPa (39 psi). Originally, concrete mounting pads were used to control vibration caused by the booster pumps. They are now replaced by airloc vibration isolation pads [5] on top of SS plates bolted to the floor.

The resonance control valves are ball valves controlled by the electronics and actuated electrically with a motor and reducing gearbox. The original valve motors have been failing due to exposure to radiation and condensing humidity. During the design phase, Empire Magnetics Inc.[6] supplied prototype motors for testing and their performance and physical size were determined to be acceptable. The motor assembly is comprised of single stack, 34 frame, 1.8-degree hybrid stepper motor, built with radiation resistant materials and rated for $2 \times 10^6$ Gray ($2 \times 10^8$ Rad) accumulated gamma ray dose. The motor alone produces more than 1.0 N m (8.8 in lbf.) of static torque at the rated amperage for the winding configuration. The motors are provided with six leads for wiring to unipolar drives. The motor and gearbox assembly delivers the minimum required torque output of 30 N-m (260 in.lbs.) at the gearbox. This motor gearbox assembly is also sealed which means it is drip resistant which could help with condensing humidity. Upgraded resonance control valves seem to have better reliability and improved performance. It is important that these valves close quickly in case of RF trips [1]. With newer motors, the time it takes to close the valves reduced from 15-20 seconds to 2-4 seconds due to reduced gear ratio.

Table 1: Characteristics of 805 MHz LLRF RCWS

| Characteristics | Values |
| --- | --- |
| Inlet Water Temperature (℃/°F) | 20/68 |
| Outlet Water Temperature (℃/°F) | 37/98 |
| Mix Tank Water Temperature (℃/°F) | 27/80 |
| Inlet Water Pressure (kPa/psi) | 690/100 |
| Outlet Water Pressure (kPa/psi) | 138-207/20-30 |
| Coolant | De-ionized water |
| De-ionization level/resistivity (Mohm*cm) | 7-10 |

The balancing valves selected are by IMS Hydronic Engineering, model number TA-BVS-240/243 [7]. They have stainless steel bodies and welded pipe connections. The valve size is the American nominal "2-inch" trade size.

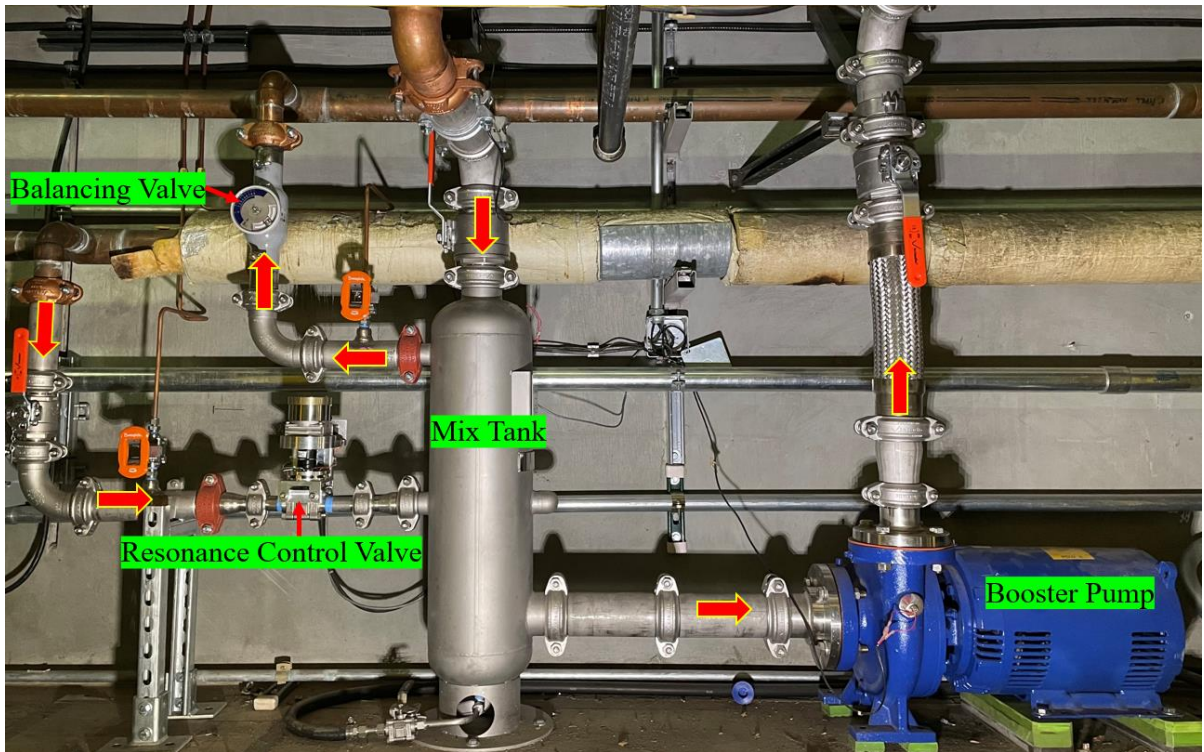

Figure 3: Upgraded 805 MHz LLRF RCWS

The flow of cool water from the main supply is regulated by the resonance control valve. To keep the pressure constant in the closed system, the balancing valve is used to regulate the flow of warm water flowing to the main return manifold. In doing so, the balancing valve maintains a constant pressure drop and ensures the flow of cooling water along the length of the module is equalized. The values of the pressure drop, and flow rate are known. The flow coefficient was calculated and used to determine the position of the valve via a table supplied by the manufacturer.

$$Kv = 36 \frac{q}{\sqrt{\Delta p}}$$

where, $Kv$ is the flow coefficient, $\Delta p$ is the pressure drop across the valve in kPa and $q$ is the design flow rate in $l/sec$. Due to limited space, the entry and exit length requirements for the balancing valve could not be met. Based on the observed performance, that criteria seems to have minimum effect on this application. Figure. 3 shows the upgraded RCWS.

## CHALLENGES ENCOUNTERED

In the design phase, the plan was to upgrade 33 of 44 total modules. However, the increased cost of materials, equipment, and fabrication during and after the pandemic necessitated the objective to be reduced to 28.

The manufacturer's staffing challenges caused extended delivery delays for the resonance control valves. Because the installation schedule was so rigid, the old resonance control valves had to be retrofitted to the upgraded systems during the extended maintenance outage in the early months of 2023. The new valves have since been delivered and are scheduled for installation in the 2024 maintenance outage.

## CONCLUSION

Out of 44 modules, 20 of them have been upgraded and 8 more will be upgraded during 2024 extended outage. We are currently looking for funding to upgrade the remaining 16 modules in the near future. Speaking generally, the progress in upgrading the LLRF RCWS seems to be improving operational efficiency. However, the full extent of the benefits will be apparent only after all 44 modules are upgraded.

## ACKNOWLEDGEMENTS

The authors would like to thank Manuel Soliz, Isaish Maldonado, Jerome DeAguero, Josh Brito, Fred Gaul and John Bernal for their hard work and contributions.

## REFERENCES

[1] M. Prokop et. al, "LANSCE- R Low Level RF Control System", Proceedings of PAC07, Albuquerque, NM, USA;

[2] Flame enterprises Inc., https://www.flamecorp.com/

[3] ASME Boiler and Pressure Vessel Code, https://www.asme.org/codes-standards/bpvc-standards

[4] Terminal Engineering and Manufacturing, https://www.terminalem.com/

[5] Airloc, https://www.airloc.com/us/home.html

[6] Empire Magnetics Inc., https://www.empiremagnetics.com/

[7] IMI Hydronic Engineering, https://www.imi-hydronic.com/en-us/product/ta-bvs-240243